\documentclass[10pt, conference, compsocconf]{IEEEtran}
\def\FIGDIR{./figures}          

\usepackage{amsmath}
\usepackage{amsthm}
\theoremstyle{definition}
\usepackage{algorithmic}
\usepackage[linesnumbered]{algorithm2e}

\usepackage[usenames,dvipsnames]{xcolor}

\usepackage{setspace}
\usepackage{comment}
\usepackage{subfigure}
\usepackage[most]{tcolorbox}
\usepackage{pifont}

\usepackage{endnotes,xspace,graphicx,fancyvrb,multirow}
\usepackage{lipsum}

\usepackage{paralist}

\usepackage{mathptmx} 

\newcommand{\ignore}[1]{}
\usepackage{fancyhdr}
\usepackage[normalem]{ulem}
\usepackage[hyphens]{url}
\usepackage{microtype}
\usepackage{flushend}
\usepackage[sort,nocompress]{cite}
\usepackage[bookmarks=true,breaklinks=true,letterpaper=true,colorlinks,linkcolor=black,citecolor=blue,urlcolor=black]{hyperref}

\usepackage{enumitem}

\usepackage{booktabs} 

\RequirePackage[nameinlink]{cleveref} 
\crefname{chapter}{Chapter}{Chapters}
\crefname{section}{Section}{Sections}
\crefname{subsection}{Subsection}{Subsections}
\crefname{equation}{Equation}{Equations}
\crefname{definition}{Definition}{Definitions}
\crefname{assumption}{Assumption}{Assumptions}
\crefname{theorem}{Theorem}{Theorems}
\crefname{figure}{Figure}{Figures}
\crefname{table}{Table}{Tables}
\let\autoref\cref 
\renewcommand{\ignore}[1]{}
\newcommand{\PCignore}[1]{}

\newcommand{\insertFigure}[2]{
    \begin{figure}[t]
\setlength{\abovecaptionskip}{-2pt}
\setlength{\belowcaptionskip}{-2pt}
        \centering
        \includegraphics[width=\linewidth]{\FIGDIR/#1.pdf}
	\vspace{-3mm}
        \caption{\small #2}
        \label{fig:#1}
    \end{figure}
}

\newcommand{\insertWideFigure}[2]{
    \begin{figure*}[h]
    \setlength{\abovecaptionskip}{-3pt}
    \setlength{\belowcaptionskip}{-4pt}
        \centering
        \includegraphics[width=\textwidth]{\FIGDIR/#1.pdf}
	    \vspace{-2mm}
        \caption{\small #2}
   	\vspace{-3mm}
        \label{fig:#1}
    \end{figure*}
}

\newcommand{\squishlist}{
 \begin{list}{$\bullet$}
  { \setlength{\itemsep}{0pt}
     \setlength{\parsep}{3pt}
     \setlength{\topsep}{3pt}
     \setlength{\partopsep}{0pt}
     \setlength{\leftmargin}{1.5em}
     \setlength{\labelwidth}{1em}
     \setlength{\labelsep}{0.5em} } }

\newcommand{\squishlisttwo}{
 \begin{list}{$\bullet$}
  { \setlength{\itemsep}{0pt}
     \setlength{\parsep}{0pt}
    \setlength{\topsep}{0pt}
    \setlength{\partopsep}{0pt}
    \setlength{\leftmargin}{2em}
    \setlength{\labelwidth}{1.5em}
    \setlength{\labelsep}{0.5em} } }

\newcommand{\squishend}{
  \end{list}  }

\newcommand{\betterparagraph}[1]{\noindent\textbf{#1.}}

\newcommand{\rev}[1]{#1}

\newcommand{\ourwork}{Herald\xspace}

\def\BibTeX{{\rm B\kern-.05em{\sc i\kern-.025em b}\kern-.08em
    T\kern-.1667em\lower.7ex\hbox{E}\kern-.125emX}}

\fancypagestyle{firstpage}{
  \fancyhf{}
  
  \fancyfoot[C]{\thepage}
}

\fancypagestyle{firstpage}{
  \fancyhf{}

  \fancyhead[C]{\normalsize{The 27th IEEE International Symposium on High-Performance Computer Architecture (HPCA 2021)}}
  \fancyfoot[C]{\thepage}
}

\author{
    \IEEEauthorblockN{Hyoukjun Kwon\IEEEauthorrefmark{1}\IEEEauthorrefmark{2}, Liangzhen Lai\IEEEauthorrefmark{2}, Michael Pellauer\IEEEauthorrefmark{3}, Tushar Krishna\IEEEauthorrefmark{1}, Yu-Hsin Chen\IEEEauthorrefmark{2}, Vikas Chandra\IEEEauthorrefmark{2}}
    \IEEEauthorblockA{\IEEEauthorrefmark{1}Georgia Insitute of Technology, \IEEEauthorrefmark{2}Facebook, \IEEEauthorrefmark{3}NVIDIA}
    \IEEEauthorblockA{
    \IEEEauthorrefmark{1}hyoukjun@gatech.edu, tushar@ece.gatech.edu
    ,\\\IEEEauthorrefmark{2}\{hyoukjunkwon, liangzhen, yhchen, vchandra\}@fb.com,     \IEEEauthorrefmark{3}mpellauer@nvidia.com
    }
}

\title{Heterogeneous Dataflow Accelerators for Multi-DNN Workloads 
}

\begin{document}
\maketitle
\thispagestyle{firstpage}
\pagestyle{plain}


\begin{abstract}
Emerging AI-enabled applications such as augmented and virtual reality (AR/VR) leverage multiple deep neural network (DNN) models for various sub-tasks such as object detection, image segmentation, eye-tracking, speech recognition, and so on.
Because of the diversity of the sub-tasks, the layers within and across the DNN models are highly heterogeneous in operation and shape.
Diverse layer operations and shapes are major challenges for a fixed dataflow accelerator (FDA) that employs a fixed dataflow strategy on a single DNN accelerator substrate since each layer prefers different dataflows (computation order and parallelization) and tile sizes.
Reconfigurable DNN accelerators (RDAs) have been proposed to adapt their dataflows to diverse layers to address the challenge.
However, the dataflow flexibility in RDAs is enabled at the cost of expensive hardware structures (switches, interconnects, controller, etc.) and requires per-layer reconfiguration, which introduces considerable energy costs.

Alternatively, this work proposes a new class of accelerators, heterogeneous dataflow accelerators (HDAs), which deploy multiple accelerator substrates (i.e., sub-accelerators), each supporting a different dataflow.
HDAs enable coarser-grained dataflow flexibility than RDAs with higher energy efficiency and lower area cost comparable to FDAs.
To exploit such benefits, hardware resource partitioning across sub-accelerators and layer execution schedule need to be carefully optimized.
Therefore, we also present \ourwork, a framework for co-optimizing hardware partitioning and layer scheduling.
Using \ourwork on a suite of AR/VR and MLPerf workloads, we identify a promising HDA architecture, Maelstrom, which demonstrates 65.3\% lower latency and 5.0\% lower energy compared to the best fixed dataflow accelerators and 22.0\% lower energy  at the cost of 20.7\% higher latency compared to a state-of-the-art reconfigurable DNN accelerator (RDA).
The results suggest that HDA is an alternative class of Pareto-optimal accelerators to RDA with strength in energy, which can be a better choice than RDAs depending on the use cases.


\end{abstract}

\section{Introduction}
\label{sec:introduction}

The success of deep learning over the past few years has led to the development of breakthrough applications such as augmented and virtual reality (AR/VR)~\cite{wu2019machine} and autonomous driving~\cite{tian2018deeptest, lee2019copy, ramanishka2018toward}.
These applications employ not one, 
but multiple deep neural networks (DNN) internally for various tasks
to collaboratively achieve state-of-the-art operational performance
\footnote{Since performance is an overloaded term, we distinguish computational (i.e., latency, throughput, etc.) and operational (i.e., the accuracy of DNN classification) performance in this paper.}
of the application.
For example, a VR application includes sub-tasks such as object detection to prevent users from conflicting with nearby obstacles, hand tracking, and hand pose estimation for user inputs, eye-tracking for foveated rendering, and so on~\cite{hazelwood2018applied, sha2019computational, kaplanyan2019deepfovea, ARVR_Michael_Abrash}.

\insertFigure{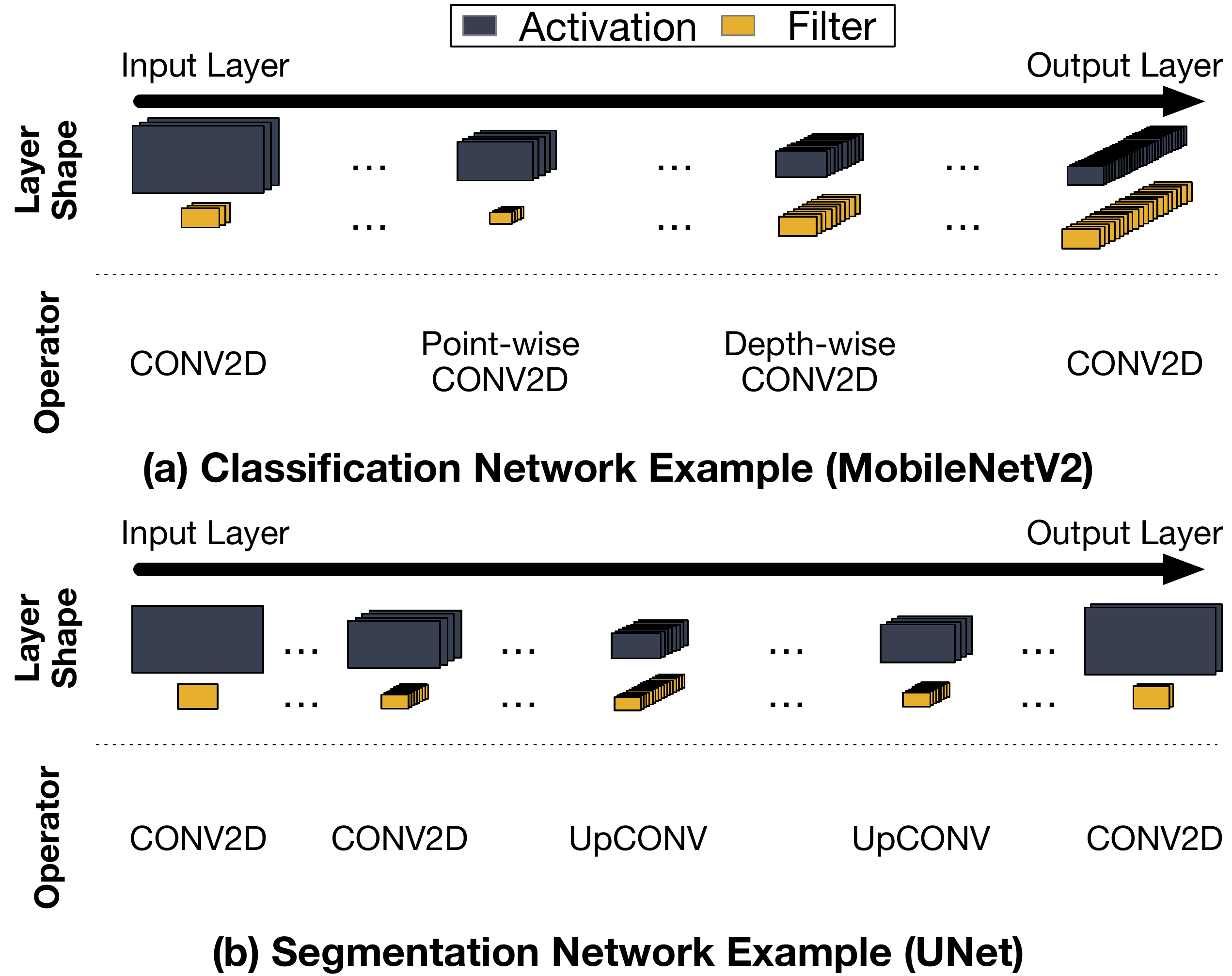}{A high level overview of layer shape (i.e., tensor shapes) of (a) classification networks such as Resnet50~\cite{Resnet} and MobileNetv2~\cite{mobilenetv2} and (b) segmentation networks such as UNet~\cite{UNet}.}

\insertFigure{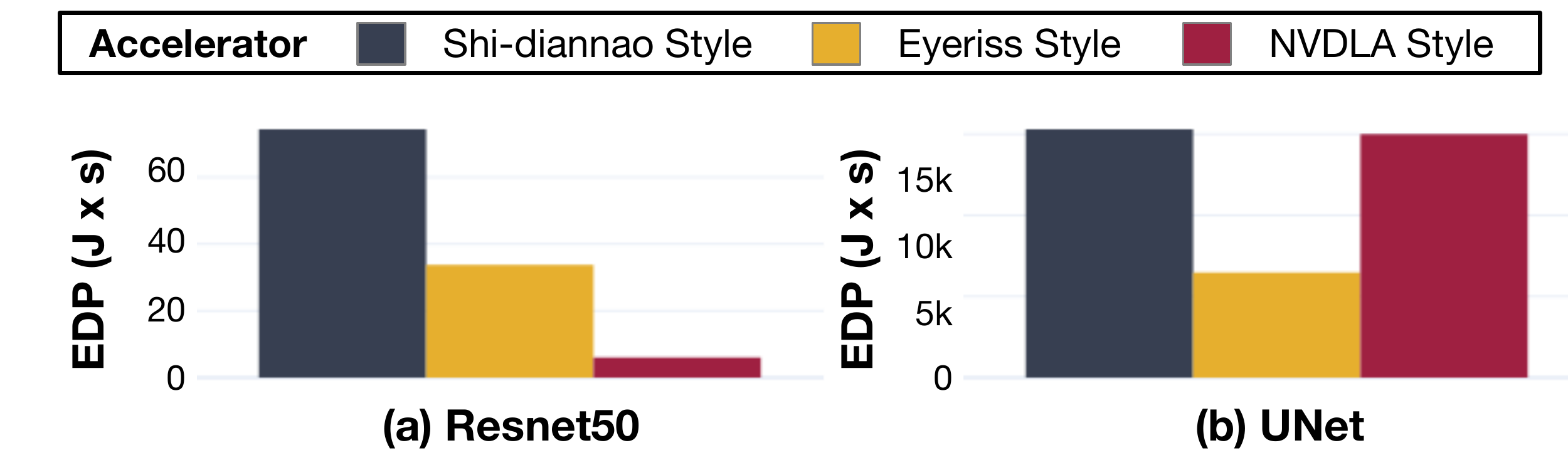}{EDP estimation of DNN accelerators with output-stationary (ShiDianNao)~\cite{du2015shidiannao}, weight-stationary (NVDLA)~\cite{nvdla}, and row-stationary (Eyeriss)~\cite{chen2016eyeriss_isca} style dataflows for running Resnet50 and UNet.
For a fair comparison, we choose 256 PEs and 32GBps NoC bandwidth for all accelerators and model them within a common framework MAESTRO~\cite{kwon2019understanding} that estimates the energy and runtime based on data reuse facilitated by the dataflow.}

\begin{table*}[]
\centering
\caption{Heterogeneity in DNN models used in AR/VR workloads~\cite{wu2019machine}. For works without model names, we name them to refer to those works in the rest of the paper. The channel-activation size ratio is an abstraction of layer shape, which refers to the number of channels divided by activation size (width or height).
CONV2D, PWCONV, DWCONV, Skip-Con, UPCONV, and Concat refer to 2D convolution, point-wise 2D convolution, depth-wise convolution, skip connection, up-scale convolution, and concatenation, respectively.
}
\label{tab:dnn_models}
\footnotesize
\begin{tabular}{|l|l|l|l|}
\hline
\multicolumn{1}{|c|}{\textbf{Task}} 
& \multicolumn{1}{c|}{\textbf{Model}} 
& \multicolumn{1}{c|}{\textbf{Channel-Activation Size Ratio}} 
& \multicolumn{1}{c|}{\textbf{Layer Operations}} \\ 
\hline
Object Detection
& MobileNetV2~\cite{mobilenetv2}
& Min: 0,013, Median: 13.714, Max: 1280
& CONV2D, PWCONV, DWCONV, Skip-Con. \\ \hline
Object Classification
& Resnet50~\cite{Resnet}
& Min: 0.013, Median: 18.286, Max: 292.571
& CONV2D, FC, Skip-Con. \\ \hline
Hand Tracking
& UNet~\cite{UNet}
& Min: 0.002, Median: 1.855, Max: 34.133
& CONV2D, FC, UPCONV, Concat. \\ \hline
Hand Pose Estimation
& Br-Q HandposeNet~\cite{madadi2017end}
& Min: 0.016, Median: 1024, Max: 1024
& CONV2D, FC \\ \hline
%
Depth Estimation
& Focal Length DepthNet~\cite{he2018learning}
& Min: 0.013, Median: 4.571, Max: 4096
& CONV2D, FC, UPCONV \\ \hline
\end{tabular}
\vspace{-2mm}
\end{table*}

\insertWideFigure{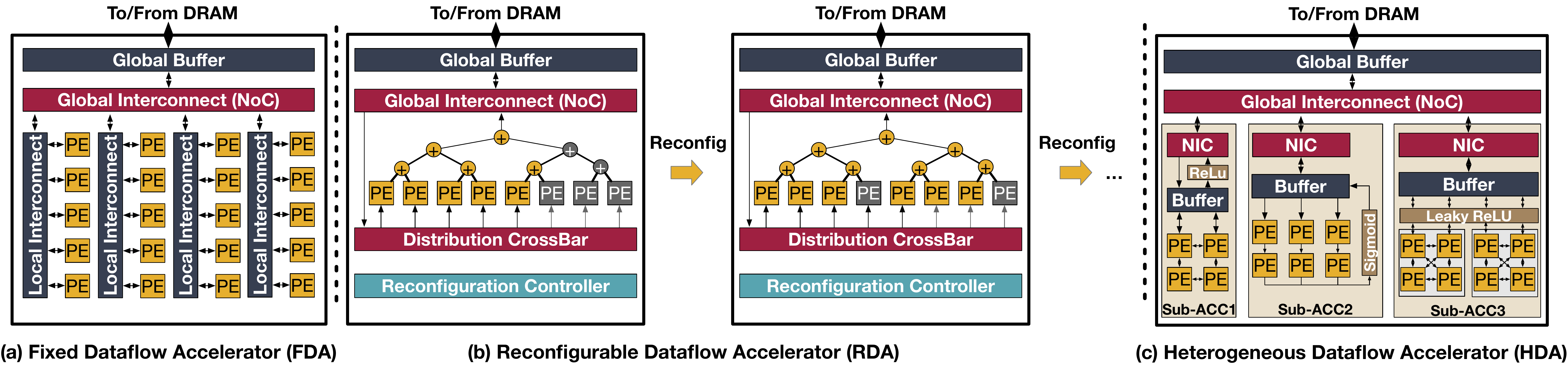}{Examples of fixed, reconfigurable, and heterogeneous dataflow accelerators.}

We list some of the DNNs used in AR/VR applications in~\autoref{tab:dnn_models}.
Despite all these DNNs being Convolutional Neural Network (CNN)-based, they exhibit high heterogeneity - both in layer shapes and operations, depending on the task.
E.g., across the layers in the example DNN models, the largest channel-activation size ratio, which is an indicator of layer shapes, is 315076$\times$ larger than the smallest one.
In terms of operations, in addition to CONV2D, these DNNs rely on operators such as depth-wise, transpose, and up-scale convolutions.
The heterogeneity is shown qualitatively in \autoref{fig:LayerShapeAndOps} and explained in~\autoref{subsec:het_multi_DNN_workloads}.

Such extreme layer heterogeneity introduces major efficiency (i.e., latency and energy efficiency) challenges to DNN accelerators since they are often over-specialized for a specific set of DNN layers, which provides them an efficiency boost in the first place.
This over-specialization is based on the 
\textit{dataflow} choice in the accelerator. 
%
We can quantitatively observe the impact of dataflow choices in~\autoref{fig:LayerPreference}, which compares energy-delay-product (EDP) of Shi-diannao~\cite{du2015shidiannao}, NVDLA~\cite{nvdla}, and Eyeriss~\cite{chen2016eyeriss_isca} style accelerators across two different DNN models.
NVDLA's dataflow exploits parallelism across input and output channels, which enables to achieve near roof-line throughput and thus low EDP for CONV2D layers with deep channels (such as those in ResNet50), as shown in~\autoref{fig:LayerPreference} (a).
However, when running CONV2D layers with a small number of channels dominant in UNet, NVDLA suffers from compute unit under-utilization, which leads to low throughput and high energy, as results in~\autoref{fig:LayerPreference}(b) show.
In contrast, a row-stationary style dataflow like Eyeriss parallelizes the computation over activation rows enables high PE utilization on such CONV2D layers. 
%
In other words, tuning an accelerator's dataflow for specific layers can lead to inefficiency across other layers.
We call these existing approaches \textit{Fixed Dataflow Accelerators} (FDAs).

The observation above is not new.
Multiple prior papers~\cite{kwon2018maeri, parashar2019timeloop, kwon2019understanding, yang2020interstellar} 
have pointed to the 
fact that the optimal dataflow and the tile size (together called a \textit{mapping}) choices
are highly dependent on the layer shape and operation, and one dataflow/mapping choice is not ideal for all layers of a model. 
Flexible accelerators, i.e., accelerators that support multiple dataflows, have been studied in the past for this challenge. 
They include 
coarse-grained reconfigurable architecture~(CGRA) style ASIC accelerators~\cite{lu2017flexflow, kwon2018maeri,chen2019eyeriss}.
We term such approaches \textit{reconfigurable dataflow accelerators}~(RDAs).
A key challenge with RDAs is that the flexibility is enabled at the cost of extra hardware components (switches and wires) that are a cause for concern for deployment under stringent energy constraints in edge, mobile, and cloud devices (e.g., MAERI~\cite{kwon2018maeri} required 11.7\% more energy, on average, compared to NVDLA-style FDA in our evaluation).
Moreover, reconfiguring for the optimal mapping for each layer~\cite{zhao2019mrna} would also add additional latency and power costs at the end of each layer.

In this work, we propose a new class of DNN accelerators called \textit{heterogeneous dataflow accelerators} (HDAs). 
HDAs provide flexibility by employing multiple sub-accelerators, each tuned for a different dataflow, within an accelerator chip.
HDAs provide two important features:
(i) \textit{dataflow flexibility}, enabled by scheduling each layer from the multiple DNN models on the most efficient sub-accelerator for each layer, as long as possible.
(ii) \textit{high utilization}, enabled by scheduling multiple layers from different models across the sub-accelerators simultaneously.

Using four example HDA architectures in the evaluation, we demonstrate that the HDA approach offers a promising mechanism for enabling dataflow flexibility similar to RDAs while staying within the area-power budget of FDAs and being more robust to workload changes.
In our evaluation, HDAs with the best EDP for each experiment provided \rev{73.6\%} lower energy-delay product (\rev{65.3\%} latency and \rev{5.0\%} energy benefits) across evaluated multi-DNN workloads, compared to the best monolithic designs we evaluate.

We summarize the contribution of this paper as follows:
\squishlist
  {\item This is the first work to propose the concept of HDAs for DNN acceleration.}
  {\item We propose a hardware and schedule co-design space exploration (DSE) algorithm that searches for (i) optimized hardware resource distribution across sub-accelerators and (ii) optimized layer execution schedules on the sub-accelerators for a given multi-DNN workload}
  {\item We codify the DSE algorithm and implement \ourwork, which can be used as by architects at \textbf{design time} by running (i) and (ii) together, or by compilers as a scheduler by running (ii) at \textbf{compile time}.}
  {\item We identify a novel HDA partitioning strategy that employs NVDLA~\cite{nvdla} and Shi-diannao~\cite{du2015shidiannao} style dataflows for unique benefits. We name this accelerator architecture \textbf{Maelstrom} and explore the scalability over edge, mobile, and cloud scenarios. On average, across three multi-DNN workloads and three scalability scenarios, Maelstrom demonstrates 65.3\% lower latency and 5.0\% lower energy compared to the best fixed dataflow accelerators, 63.1\% lower latency and 4.1\% lower energy compared to the homogeneous multi-DNN~\cite{baek2020multi}-style accelerators, and 20.7\% higher latency and 22.0\% lower energy compared to an exiting reconfigurable accelerator~\cite{kwon2018maeri}.}
\squishend

\section{Background and Motivation}
\label{sec:background}

\subsection{Heterogeneous Multi-DNN Workloads}
\label{subsec:het_multi_DNN_workloads}

To achieve high-quality (classification, prediction, etc.) results, many applications now employ DNNs as their backbone to perform tasks like face recognition~\cite{schroff2015facenet}, image segmentation~\cite{UNet, he2017mask}, depth estimation~\cite{he2018learning}, and so on.
Combining DNNs for such tasks, emerging applications such as AR/VR implement complex functionalities, which lead to multi-DNN workloads~\cite{wu2019machine}.
These sub-task DNNs are significantly diverse, as shown in~\autoref{tab:dnn_models}.
The diversity of models naturally leads to high variations in layer (1) shape and (2) operations, which constructs heterogeneous multi-DNN workloads.

\subsubsection{Layer Shape}
\label{subsubsec:layer_shape}

Classification networks such as Resnet~\cite{Resnet} or MobileNetV2~\cite{mobilenetv2} gradually reduce the resolution of activation because their goal is to extract a classification vector where each entry represents the probability of each class.
Also, classification networks increase the number of channels to exploit as many features as possible for accurate classification.
Therefore, layers in classification networks have high-resolution activation and shallow channels in early layers and low-resolution activation and deep channels in late layers, as illustrated in~\autoref{fig:LayerShapeAndOps} (a).

In contrast, segmentation networks such as UNet~\cite{UNet} need to restore the original resolution of activation because their goal is to generate masks over target objects in the input image.
However, segmentation networks still need to extract as many features as those in classification networks for high accuracy.
Therefore, segmentation networks first follow the same trend as classification networks until the mid-layer.
Afterward, segmentation networks reduce the number of channels and gradually restore the resolution of activation using up-scaling operators such as up-scale convolution or transposed convolution.
As a result, layer shapes in segmentation networks follow the trend illustrated in~\autoref{fig:LayerShapeAndOps} (b).

\subsubsection{Layer Operation}
\label{subsubsec:layer_operation}

As listed in the layer operation column of~\autoref{tab:dnn_models}, layer operations in heterogeneous multi-DNN workloads are diverse.
For example, MobileNetV2 performs depth-wise separable convolution~\cite{mobilenetv2}, which consists of two point-wise convolutions and a depth-wise convolution. 

Based on the shape and operation, each layer prefers different dataflow styles and hardware~\cite{kwon2019understanding}, which makes such workloads challenging for fixed dataflow accelerators (FDAs).
We clarify the definition of dataflow and mapping and discuss why each layer prefers different dataflows next.

\subsection{Dataflow and Mapping}
\label{subsec:dataflow}

\insertFigure{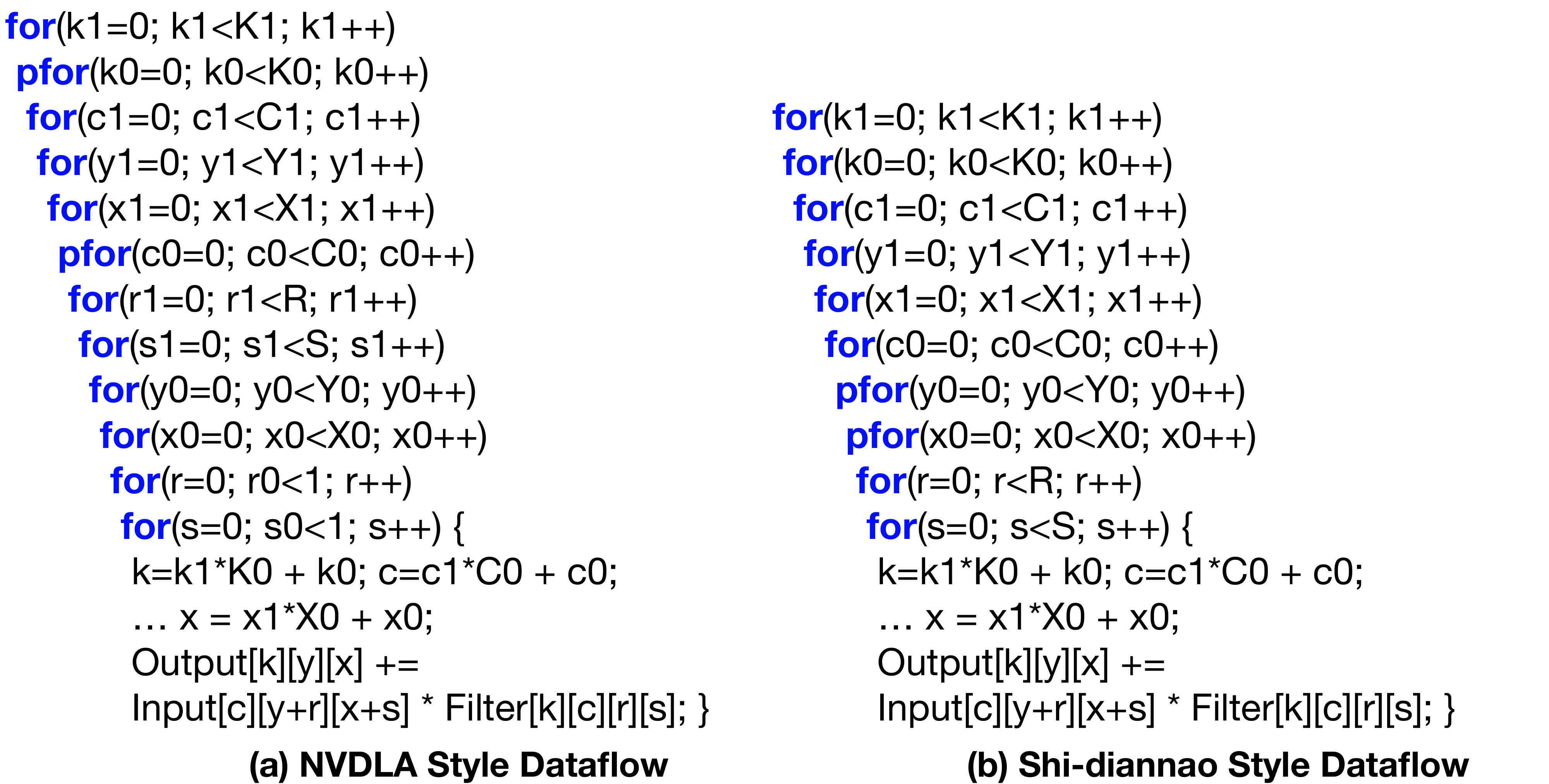}{Loop nest representation of dataflows from recent accelerators~\cite{chen2016eyeriss_isca, du2015shidiannao}.
K and C refer to output and input channels, Y and X refer to input row and column, and R and S refer to filter row and column, respectively.
Numbers after loop variables indicate tile levels, and pfor refers to a parallel for loop.
We omit edge case handling for simplicity.
} 

We collectively refer to loop ordering and spatial unrolling (or partitioning) as dataflow~\cite{chen2016eyeriss_isca, yang2020interstellar}.
Dataflows are often represented in a loop-nest form~\cite{chen2016eyeriss_jssc}, as shown in~\autoref{fig:Dataflows}, loop nest with loop bounds are unfilled.
From a base loop nest without any loop transformation, a series of loop interchange and parallelization modifies how we compute DNN operations while preserving what we compute.
The loop nests in~\autoref{fig:Dataflows} are results of such loop transformations.
By providing valid loop bounds to the representation (i.e., loop blocking factor), we obtain ``mapping," which indicates an instance of dataflow, which contains full information to map a DNN operation on an accelerator~\cite{kwon2020maestro}.
In DNN accelerators, the mapping dictates the latency and energy consumption because it determines the number of buffer accesses, degree of parallelization (mapping utilization of PEs), buffer size requirements, and so on~\cite{chen2016eyeriss_isca, lu2017flexflow, parashar2019timeloop, kwon2019understanding, yang2020interstellar}.

\insertWideFigure{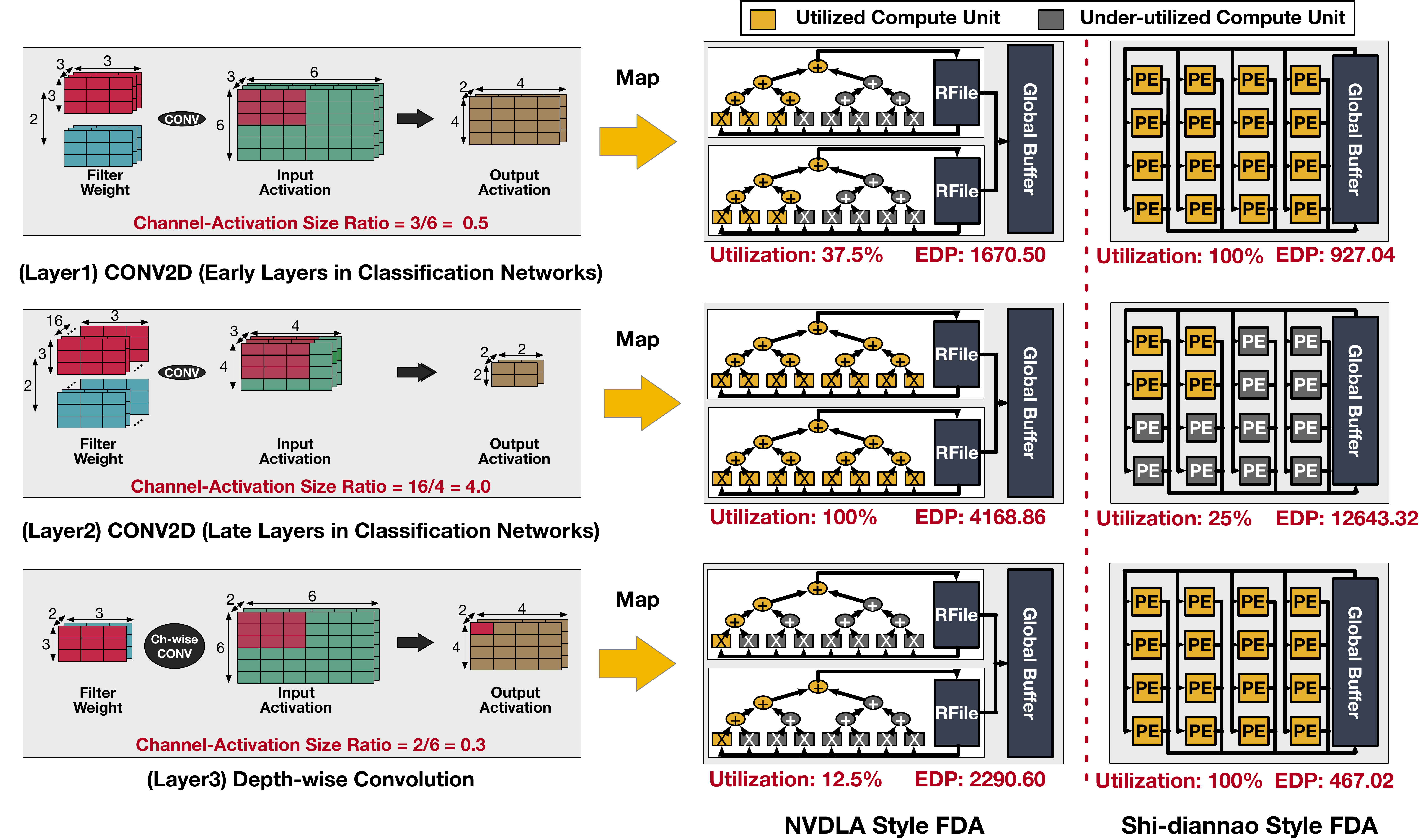}{The impact of dataflow styles on efficiency. We show three example layer execution scenarios on NVDLA and Shi-diannao style FDAs. The utilization refers to the mapping utilization of compute units, which refers to the ratio of the number of PEs a mapping utilizes and the total number of PEs in an accelerator.
}

When constructing a mapping from a dataflow, the set of valid loop bounds (loop blocking factors) are constrained by the sizes of each layer dimension (i.e., out-of-bound) and layer operation (e.g., depth-wise convolution does not accumulate partial sums across input channels unlike CONV2D).
Such constraints to loop bounds from layer shape and operation determine a set of available mappings from each dataflow, which appears as the preferences to layers.
To further understand such an aspect, we use two example FDAs with NVDLA and Shi-diannao dataflow styles and three example layers shown in~\autoref{fig:background_dataflow_impact}.
For simplicity, we select the minimum loop blocking factor to construct mappings for each dataflow.

Those two example accelerators have distinct approaches to compute MAC operations in DNNs.
As illustrated in~\autoref{fig:background_dataflow_impact}, a Shi-diannao style accelerator parallelizes the computation over output activation row and column dimensions using an output-stationary style dataflow, which exploits output and convolutional reuse.
Unlike the example Shi-dianao style accelerator, the example NVDLA style accelerator in~\autoref{fig:background_dataflow_impact} parallelizes the computation over input and output channels using a weight-stationary style dataflow, which exploits filter weight reuse.
Such differences in parallelization strategies of the example dataflows result in dramatically different mapping utilization of compute units, as shown in~\autoref{fig:background_dataflow_impact}.
We use three example layers presented in~\autoref{fig:background_dataflow_impact} to show the impact of mappings.
Layer 1 and 2 are CONV2D operations with the aspect ratio of early and late layers in classification network introduced in~\autoref{fig:LayerShapeAndOps} (a), respectively.
Layer 3 is a depth-wise CONV2D operation with the same layer size as Layer 1.

Based on the parallelization strategies of each example accelerator and layer sizes, we can observe dramatically different PE utilization as shown in~\autoref{fig:background_dataflow_impact}.
We use MAESTRO~\cite{kwon2019understanding} cost model for DNN accelerators to estimate the latency and energy and compute energy-delay product (EDP) as one of the indicators of overall efficiency, as shown in~\autoref{fig:background_dataflow_impact}.
In the combination of the differences in utilization and data reuse strategies, two example accelerators result in dramatically different EDPs, which implies distinct preferences of two example accelerators to the layers.
In addition to the mapping utilization, each of the example mappings has dramatically different memory/network-on-chip(NoC) bandwidth requirements, buffer size requirements, and so on, which also varies based on the layer shape and operations in a different degree~\cite{kwon2019understanding}.

Therefore, no single dataflow style is good for all the layers, and we need to optimize the dataflow for each layer in target workloads to maximize the efficiency of an accelerator.
However, when the target workload is heterogeneous, the common practice that optimizes the dataflow for the average case of the workload can result in a consistently inefficient mapping for all the layers in the workload, which is one of the major challenges for DNN acceleration for emerging applications with multiple DNN models.
We discuss available accelerator options to deal with such a problem with a general introduction to DNN accelerators next.

\section{Heterogeneous Dataflow Accelerators (HDAs)}
\label{sec:hda_idea}

We propose heterogeneous dataflow accelerators (HDAs) that deploy multiple FDA instances within a chip, each running a different dataflow, as illustrated in~\autoref{fig:HetAcc} (c).
This approach eliminates extra hardware costs for reconfigurability and enables dataflow flexibility by enabling the selection of appropriate sub-accelerators for each layer.
To maximize the efficiency of computation, HDAs by default assign layers to a sub-accelerator with the most preferred dataflow style for each layer.
However, to exploit more benefits by HDAs under the efficiency drop from smaller sub-accelerators than full FDAs or RDAs, HDAs require new design considerations that did not exist in FDAs or RDAs.
We first discuss such design considerations and formally define the HDA architecture based on them.
Then, we highlight aspects of HDAs that provide benefits and challenges.

\subsection{Design Considerations and Definition of HDA}
\label{subsec:design_considerations}

To design an HDA, we need to (1) select dataflows for sub-accelerators, (2) determine how to partition existing hardware resources across sub-accelerators, and (3) find a legal layer execution schedule that satisfies layer dependence and memory size constraints.
We discuss those three design considerations.

\betterparagraph{Dataflow Selection for Sub-accelerators} 
As we show in~\autoref{sec:evaluation}, the dataflow styles to build sub-accelerators of an HDA is crucial for overall efficiency.
To maximize the benefits from dataflow flexibility, the dataflow styles of sub-accelerators need to be sufficiently different so that the resulting HDA can adapt to different layers with diverse shapes and operations.

\betterparagraph{Hardware Resource Partitioning}
As a PE partitioning example in~\autoref{fig:PE_Partitioning} shows, evenly distributed hardware resources across sub-accelerators often lead to sub-optimal HDA design points.
This shows that resource partitioning is a non-trivial optimization problem.
Also, the optimal distribution depends on workloads and selected dataflows, which makes determining hardware resource distribution further challenging.

\betterparagraph{Layer Scheduling}
Because HDAs include multiple accelerator instances, we need to determine the layer execution schedule across sub-accelerators, which is important for exploiting layer parallelism.
A scheduler must check if generated schedules are valid in terms of layer dependence and memory constraints.
In addition to that, the scheduler needs to optimize overall latency and energy, not those of each layer (i.e., global optimization for the entire HDA, not local optimization for each sub-accelerator).
Designing a scheduler satisfy all of the aforementioned requirements is challenging, and boosting the scheduler's speed to facilitate fast hardware and schedule co-design space exploration (DSE) of HDAs is another challenge.
Based on the design considerations, we formally define the HDA architecture as follows:

\newtheorem{definition}{Definition}
\begin{definition}
\textbf{HDA Architecture} \\
For given $N_{d}$ dataflow styles, D = \{$\delta_1, \delta_2, ...,\delta_{N_{d}}$ \}, total number of PEs, $N_{PE}$, total global NoC bandwidth $BW_{G}$, an HDA architecture $H$ is defined as follows: \\
$H$ = \{($\delta_i$, $N_i$, $BW_i$) $|$ $ 1 \leq i \leq n$ $\land$  $\sum N_i = N_{d}$ $\land$ $\sum BW_{i} = BW_{G}$ \}
\label{def:HDA}
\end{definition}
\vspace{-4mm}
The definition specifies the dataflow styles for each sub-accelerator, PE and bandwidth partitioning across sub-accelerators, and the total number of sub-accelerators, which fully specifies all the HDA-specific design parameters.

\subsection{Benefits of HDAs}
\label{subsec:benefits}

When optimized properly, HDAs provide latency and energy benefits based on the following three aspects.

\betterparagraph{Selective Scheduling} Because each layer prefers different dataflow and hardware, running each layer on its most preferred sub-accelerator in an HDA is an effective solution to maximize overall efficiency.

\betterparagraph{Layer Parallelism} Unlike most FDAs and RDAs that run one layer and another, HDAs can simultaneously run multiple layers of different models.
By this approach, an HDA can overlap the latency of multiple models, which leads to latency hiding among DNN models reducing overall latency.

\betterparagraph{Low Hardware Cost for Dataflow Flexibility} Because HDA employs FDA style sub-accelerators, HDAs do not involve the costs for reconfigurability like RDAs.

\subsection{Challenges for HDAs}
\label{subsec:challenges}

As we discussed in~\autoref{subsec:design_considerations}, optimizing HDA requires various considerations at once, which makes the HDA design challenging.
We discuss three aspects of the challenges.

\betterparagraph{Reduced Parallelism for Each Layer} Given the same number of PEs for an FDA and an HDA, sub-accelerators in the HDA have smaller numbers of PEs than the FDA since hardware resources need to be distributed (or partitioned) across sub-accelerators.
Therefore, the maximum degree of parallelism for each sub-accelerator decreases compared to an FDA or an RDA with the same number of PEs in total.
A smaller number of PEs and less available parallelism can lead to not only higher latency but also higher energy consumption since the amount of spatial data reuse (i.e., multicast factor) also decreases~\cite{kwon2019understanding}.

\betterparagraph{Shared Memory and NoC Bandwidth}
Because multiple sub-accelerators share a global scratchpad memory and global NoC, those resources either need to be time-multiplexed or hard-partitioned across sub-accelerators.
Like the smaller number of PEs in sub-accelerators of an HDA compared to FDAs or RDAs can lead to potential inefficiencies, lower memory and NoC bandwidth can also lead to higher latency.
To mitigate this, exploiting as much layer parallelism across sub-accelerators as possible is a key, which motivates a good scheduler.

\betterparagraph{Scheduling under Memory and Dependence Constraints to Minimize Dark Silicon} 
One of the important aspects of HDAs to enhance overall efficiency is layer parallelism to exploit as many compute units in sub-accelerators as possible, which requires a good layer scheduler.
In addition to maximizing the utilization, the layer scheduler needs to consider constraints from layer dependence and global memory size.
The scheduler needs to assign layers on the most preferred sub-accelerator to exploit the benefits of flexible dataflow.
However, such a simple greedy method that seeks a locally optimal schedule can result in a globally sub-optimal schedule, which is another challenge for the scheduler. 

To enable more benefits over discussed challenges, HDA needs to be carefully optimized.
Because many design considerations need to be considered at once, and determining one affects the optimal combinations of others, we need a systematic approach to optimize HDAs.
Therefore, we develop a hardware and schedule co-design space exploration (DSE) algorithm for HDAs that co-optimize all the design considerations in hardware and schedule.
We codify the algorithm and implement \ourwork, an HDA optimization framework, which automates HDA design tailored for user-specified target models and outputs estimated latency and energy using the co-optimized design.
We discuss the DSE algorithm and its implementation, \ourwork, next.

%

\section{Design Space Exploration Algorithm for HDAs}

We discuss the HDA co-design space exploration (DSE) algorithm that co-optimizes hardware resource partitioning across sub-accelerators and layer execution schedule.
We first discuss the execution model of HDA and the latency/energy estimation methodology we use.
Then we discuss the implementation of the DSE algorithm, \ourwork.

\subsection{Execution Model}
\label{subsec:exec_model}

We target layer granularity execution on each sub-accelerator of HDAs to (1) exploit significantly different dataflow preference of layers we discussed in~\autoref{subsec:dataflow} and (2) more fine-grained scheduling (e.g., parallelize computation tiles of one layer across multiple sub-accelerators) results in high control, synchronization, and scheduling overhead.
We assume the following execution steps of accelerators in \ourwork.

\begin{enumerate}[topsep=0pt,itemsep=-0.5ex,partopsep=1ex,parsep=1ex]
    {\item Fetch global buffer level filter weight tile from DRAM to a global buffer.}
    {\item Distribute sub-accelerator level filter weight tiles to sub-accelerators based layer execution schedule.}
    {\item Fetch global buffer level activation tile from DRAM to the global buffer.}
    {\item Stream sub-accelerator level activation tiles into their corresponding sub-accelerators based on layer execution schedule.}
    {\item Store streamed-out output activation from each sub-accelerator to the global buffer.}
    {\item Overlapping the computation and data fetch from DRAM, pre-fetch next activation and filter tiles (double buffering).
    during sub-accelerators compute output activation, fetch next filter values from DRAM and send the filter values to the next accelerator (assumes double-buffering).}
    {\item When a sub-accelerator finishes executing a layer, stream output activation stored in the global buffer as input activation of the next layer.}
    {\item Repeat above processes until processing all the layers of all the models.}
\end{enumerate}

For steps 3 and 6, activation is stored in DRAM and loaded in a tiled manner specified by the mapping in the target sub-accelerator if the buffer size is not sufficient to store the entire activation.
When output activation is committed to the global buffer, \ourwork by default assumes a rearrange buffer that adjusts the data layout for the next layer if it runs on another sub-accelerator with a different dataflow style.
In the evaluation, we select dataflows that have the same inner-loop order so that we can maintain the same data layout, which eliminates sub-accelerator context change overheads from different data layouts.
For the data layout and miscellaneous context change overheads, \ourwork also provides an option to specify the latency and energy penalties for them.

\subsection{Latency and Energy Estimation}
\label{subsec:maestro}

To guide the design space exploration, we need a cost model that estimates the latency and energy for a given HDA, a workload, and a schedule.
We use MAESTRO as a base cost model, which is a validated cost model for monolithic DNN accelerators (i.e., FDAs and RDAs) with any dataflow, which reported 96.1\% accuracy against RTL simulation~\cite{kwon2018maeri} and real processing time measured on a chip~\cite{chen2016eyeriss_jssc}.
We extend MAESTRO~\cite{kwon2019understanding, kwon2020maestro} cost model to support multi-DNN sub-accelerator environments, including HDAs.
%
%
%
%

The implementation of the DSE algorithm, \ourwork, models the memory requirement for the global buffer and data movement from/to the global buffer to/from sub-accelerator buffers.
The modeling method follows the same methodology proposed by MAESTRO, which identifies the amount of reuse and computing activity counts based on them (for energy) and communication/computation delay considering reuse (for latency).
In addition to the same analytic equations, \ourwork considers the layer execution schedule generated by our scheduler discussed in~\autoref{subsec:scheduling} by modeling non-synchronized execution of sub-accelerators (i.e., each sub-accelerator start processing a layer as soon as input data are available).
For estimating the latency and energy of each sub-accelerator run, we exploit the original MAESTRO cost model.

Next, we discuss two major steps in our DSE algorithm for HDAs: hardware resource partitioning and layer scheduling.

\subsection{Hardware Resource Partitioning Optimization}
\label{subsec:accelerator_dse}

\insertFigure{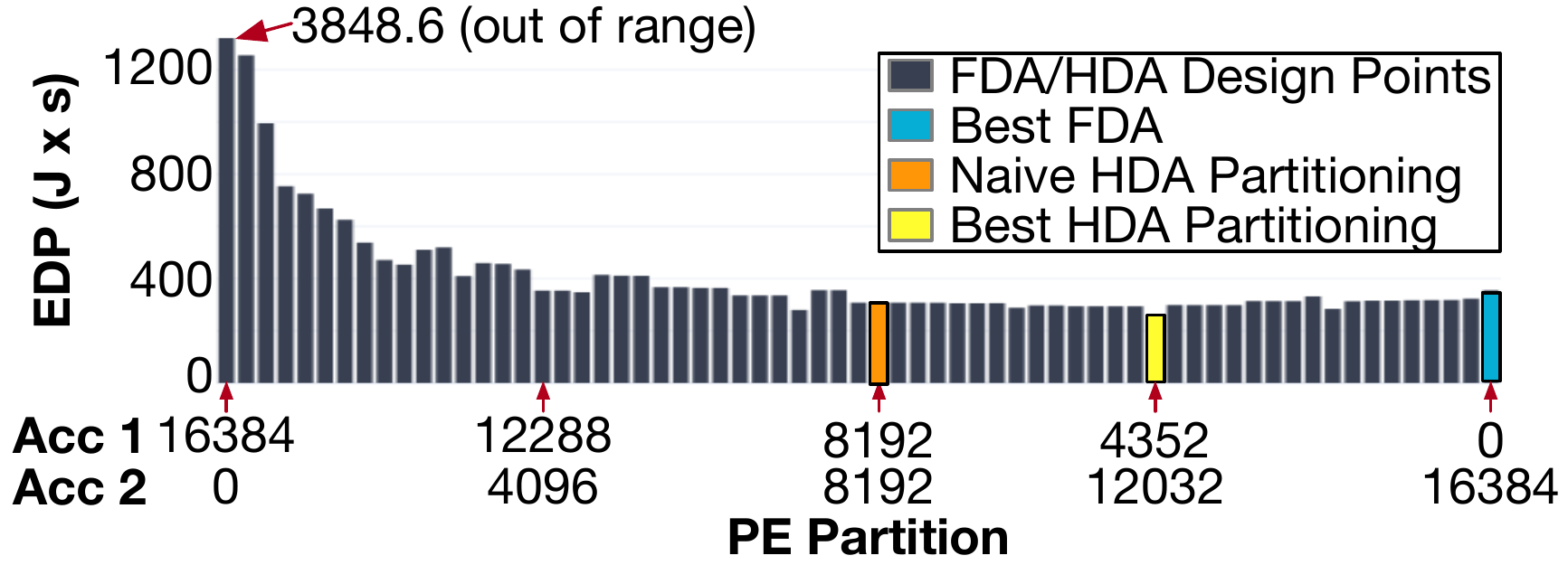}{The impact of PE partitioning upon a cloud accelerator listed in~\autoref{tab:eval_accelerators_class} with two sub-accelerators (ACC1: Shi-diannao style, ACC2: NVDLA style) with naive bandwidth partitioning. We use AR/VR-A workload presented in~\autoref{sec:evaluation}. The left- and right-most represents ACC1 and ACC2 FDA designs.
}


Unlike FDAs and RDAs fully exploit hardware resources and implement a monolithic accelerator substrate, HDAs need to distribute such resources across sub-accelerators.
However, evenly distributing those resources does not yield the most optimal HDAs because each sub-accelerator's dataflow style has different bandwidth requirements~\cite{ guirado2019understanding} and efficiency for a given number of PEs~\cite{kwon2019understanding}.

To quantitatively show the non-trivial design space of hardware resource partitioning, we show an example in~\autoref{subsec:scheduling} that shows the impact of PE partitioning over two sub-accelerators in a 16K-PE-HDA.
We show the pure impact of PE partitioning using naive bandwidth partitioning (128/128 GBps) with layer execution schedules generated by \ourwork's scheduler we discuss in~\autoref{subsec:scheduling}.
We observe that evenly partitioned PEs (8K/8K) result in a sub-optimal design point, with 17\% higher EDP than the optimal PE partition.
Therefore, we explore the PE partitioning space to identify the optimal partitioning strategy.
In addition to PEs, we also explore the global memory/NoC bandwidth partitioning, which models the hard-partitioning of global NoC wires dedicating partitioned wires for each sub-accelerator.
Such an approach is to (1) provide the right amount of bandwidth for each sub-accelerator based on the fact that the bandwidth requirement depends on dataflow choices~\cite{kwon2019understanding, guirado2019understanding} and (2) minimize hardware costs for fully flexible (i.e., all-to-all) global NoC.

To explore the partitioning choices for given dataflow styles, we implement an algorithm that explores HDA architectures defined in~\autoref{def:HDA}.
The DSE algorithm, by default, performs an exhaustive search based on user-specified search granularity.
However, the DSE algorithm also supports binary sampling or random search, which significantly reduces the search time at the cost of possible loss of globally optimal design points.

\subsection{Layer Execution Schedule Optimization}
\label{subsec:scheduling}

\insertWideFigure{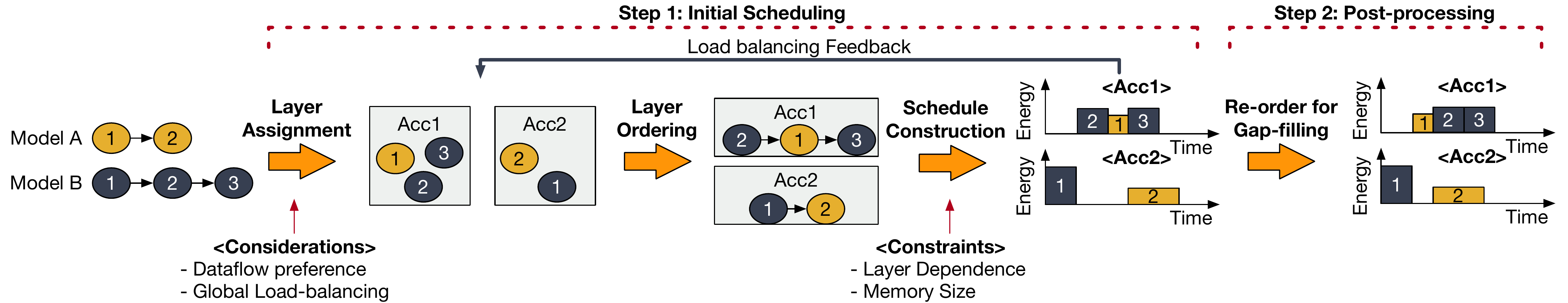}{An overview of layer scheduling algorithm of \ourwork. Circled numbers represent layers in each model.}

\insertFigure{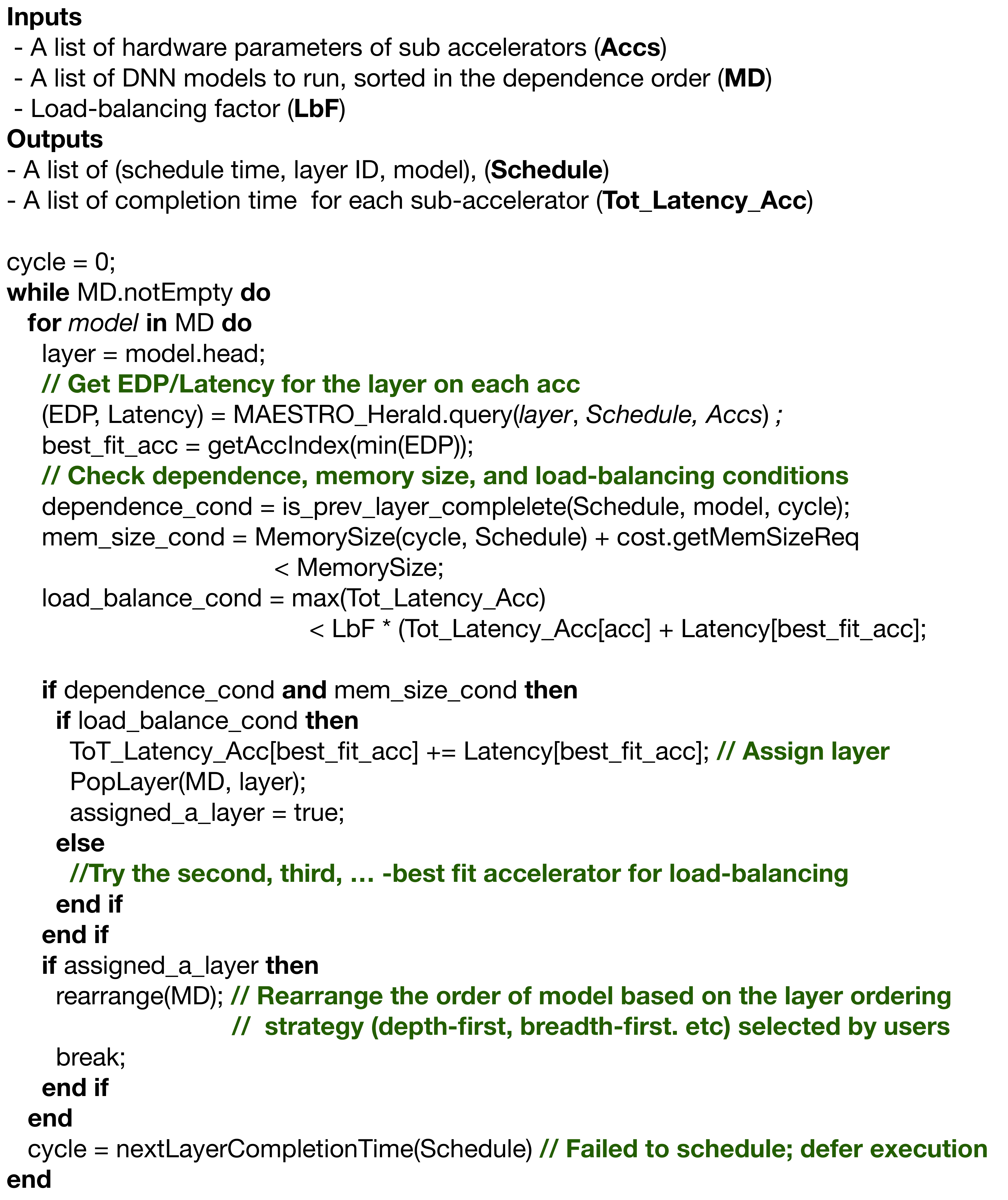}{Layer assignment and ordering algorithm.}

The main challenge for a scheduling algorithm for HDA is the massive space of schedules.
For example, $2.54 \times 10^{21}$ possible layer execution schedules exist for AR/VR-A workload in ~\autoref{tab:eval_workloads} even if we only consider permutation of the layers on a single accelerator.
To deal with such a large search space, we develop a set of heuristics that exploit the characteristics of DNN workloads to reduce the scheduling overhead.
Combining the heuristics, we implement our scheduling algorithm presented in~\autoref{fig:Scheduling_Overview}.
We discuss the heuristics we employ.

\betterparagraph{Dataflow preference-based layer assignment on sub-accelerators}
Exploiting the preferences toward dataflows, our scheduler, by default, assigns each layer on a sub-accelerator with the most-preferred dataflow.
Our scheduler implements such greedy methods, and users can select the metric (e.g., EDP, energy, latency, and so on) for them.

Because greedy methods often result in a locally optimal schedule (i.e., optimal for each layer) and miss globally optimal one, our scheduler implements a feedback loop \rev{for global load-balancing} from the initial schedule constructed after layer ordering.
When the scheduler detects an unbalanced load across sub-accelerators, the scheduler explores alternative layer assignment that reduces overall costs (EDP, energy, latency, and so on, specified by users).
Users can specify the maximum allowed load-unbalancing factor, the largest latency across sub-accelerators divided by the smallest one.
Our scheduler detects an unbalanced load based on the factor.

\betterparagraph{Heuristic-based Initial Layer Ordering}
Our scheduler exploits the characteristics of layer dependence of multi-DNN workloads: layers have mostly (i) linear dependence chain within each model, and (ii) layers are independent across models. 
Exploiting (i), our scheduler implements a depth-first layer ordering algorithm, which schedules all the layers in a DNN model first and moves on to another.
Exploiting (ii), our scheduler implements a breadth-first layer ordering algorithm, which interleaves the layer execution of each DNN model.
Those two layer ordering algorithms do not provide the optimal layer order, but they enable to quickly construct a valid initial layer execution order.
We describe the layer assignment and ordering algorithm in~\autoref{fig:LayerAssignment_Algorithm}.

\insertFigure{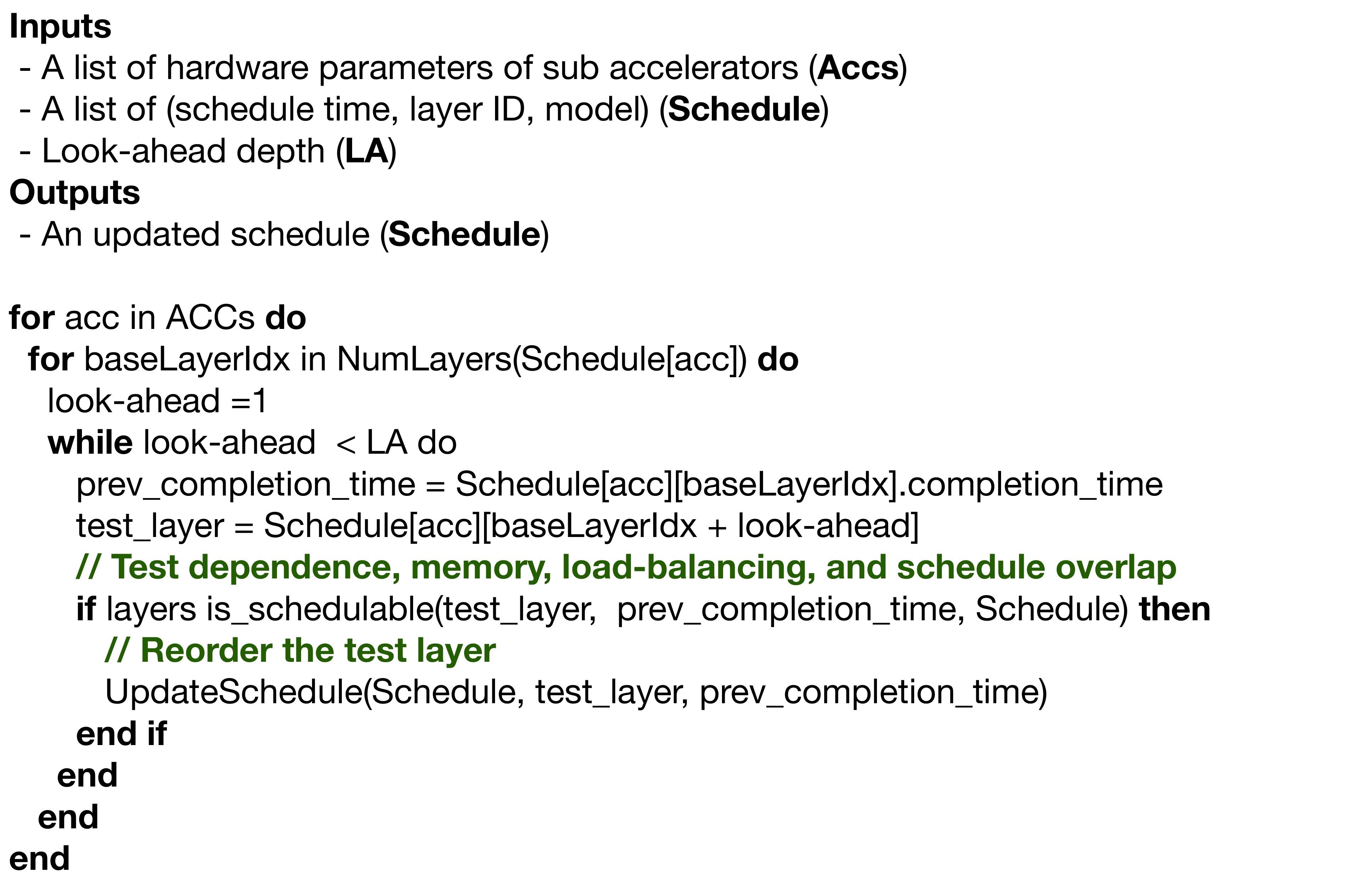}{Post-processing algorithm that removes idle time based on bad layer execution order.}

\betterparagraph{Eliminating Redundant Idle Time In Initial Schedules via Post-processing}
The initial schedule based on simple depth-first or breadth-first layer ordering often has unnecessary idle time based on bad layer execution order.
The post-processing algorithm fixes such inefficiencies with O(mn) complexity (m: total number of DNN models, n: total number of layers).
For each scheduled layer X, the algorithm search for a layer Y scheduled later than layer X that can be scheduled at the completion time of layer X.
If a layer Y is found, the algorithm re-order the layers to have layer Y right after layer X.

To identify layer Y, the post-processing algorithm searches for m (m: total number of DNN models) layers, which are head layers of each model at the completion time of layer X. 
This approach not only reduces the complexity but also ensure the layer dependence is not violated after re-ordering.
Note that this approach is only possible on valid layer schedules based on valid layer order, which is obtained quickly by simple heuristics our scheduler exploits.
We describe the post-processing algorithm in~\autoref{fig:Post_processing}.

\subsection{\ourwork: An Implementation of the DSE algorithm}

\insertWideFigure{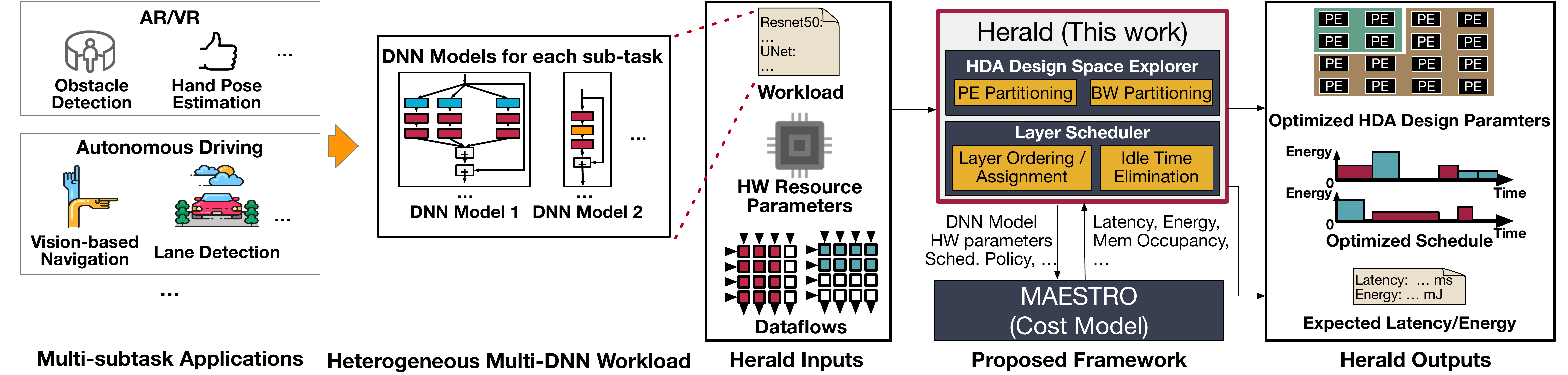}{Multi-DNN workloads from multi-subtask applications, which motivates heterogeneous DNN accelerators (HDAs). Targeting such workloads, we design an HDA optimization framework, \ourwork.}

We codify the DSE algorithms we discussed in~\autoref{subsec:accelerator_dse} and~\autoref{subsec:scheduling} and develop \ourwork.
As illustrated in~\autoref{fig:HeraldOverview}, \ourwork receives user-selected dataflow styles for sub-accelerators and co-optimizes hardware resource distribution and layer execution schedule.
\ourwork reports optimized PE and global NoC bandwidth partitioning with an optimized layer execution schedule for the partitioned sub-accelerators as outputs.
\ourwork also reports estimated total latency and energy based on MAESTRO cost model~\cite{kwon2019understanding} we extend for HDA use cases.
Using \ourwork, we evaluate four example HDA architectures and identify one HDA architecture based on NVDLA~\cite{nvdla} and Shi-diannao~\cite{du2015shidiannao} dataflow styles that provide Pareto-optimal design points among various FDAs, RDAs, and scaled-out multi-DNN FDAs (SM-FDA)~\cite{baek2020multi}.
We discuss the evaluation we perform using \ourwork next.

\section{Evaluations}
\label{sec:evaluation}

To show the potential of HDAs, we evaluate four HDA designs with layer execution schedules generated by \ourwork using three workloads listed in ~\autoref{tab:eval_workloads}.

\subsection{Evaluation Settings}
\label{subsec:eval_environment}

\begin{table}[]
\vspace{-4mm} 
\centering
\caption{Heterogeneous multi-DNN workloads used for the evaluation. We model AR/VR workloads using models listed in~\autoref{tab:dnn_models} and MLPerf~\cite{mattson2020mlperf}. 
}
\label{tab:eval_workloads}
\footnotesize
\begin{tabular}{|c|c|c|}
\hline
\textbf{Workload} 
& \textbf{Model} 
& \textbf{\# of batches} \\ 
\hline
\multirow{3}{*}{AR/VR-A}
& Resnet50 & 2 \\ \cline{2-3} 
& Unet & 4 \\ \cline{2-3} 
& MobileNetV2 & 4 \\
\hline
\multirow{5}{*}{AR/VR-B}
& Resnet50 & 2 \\ \cline{2-3} 
& Unet & 2 \\ \cline{2-3} 
& MobileNetV2 & 4 \\ \cline{2-3} 
& BR-Q Handpose & 2 \\ \cline{2-3} 
& Focal Length DepthNet & 2 \\ 
\hline
\multirow{5}{*}{\rev{MLPerf}}
& Resnet50 & \rev{1 (and 8 for batch size study)} \\ \cline{2-3} 
& MobileNetV1 & \rev{1 (and 8 for batch size study)} \\ \cline{2-3}
& SSD-Resnet34 & \rev{1 (and 8 for batch size study)} \\ \cline{2-3} 
& SSD-MobileNetV1 & \rev{1 (and 8 for batch size study)} \\ \cline{2-3} 
& \rev{GNMT (RNN)} & \rev{1 (and 8 for batch size study)} \\ 
\hline
%
%
%
\end{tabular}
\end{table}

\betterparagraph{Workloads}
Based on AR/VR-motivated DNN models listed in~\autoref{tab:dnn_models}, we construct evaluation AR/VR workloads as listed in~\autoref{tab:eval_workloads}.
For each DNN model, we assign different numbers of batches to model different target processing rate of each sub-task.
In addition to the AR/VR workloads, we also evaluate ML-perf inference workload modeling multi-stream.

\betterparagraph{Dataflow}
We combine two and three distinct dataflow styles from recent DNN accelerators(Shi-diannao~\cite{du2015shidiannao}, NVDLA~\cite{nvdla}), and Eyeriss~\cite{chen2016eyeriss_isca}.
The selection of dataflow style is based on their distinct parallelization and data reuse strategies to maximize synergy.
For example, Shi-diannao's dataflow parallelizes computations across output activation rows and columns and performs temporal accumulation of partial sums.
In contrast, NVDLA's dataflow parallelizes computations across input and output channels and performs spatial accumulation of partial sums across input channels.

Combining dataflows with such different characteristics provide more dataflow flexibility than combining similar or the same dataflows.
When we combine the same dataflow, we construct scaled-out multi-FDAs~\cite{baek2020multi}, which we also evaluate in~\autoref{subsec:eval_results}.

\betterparagraph{Cost Estimation}
As we discussed in~\autoref{subsec:maestro}, we extend MAESTRO for the latency and energy estimation.

\begin{table}[]
\centering
\caption{Evaluated accelerator styles.}
\label{tab:eval_accelerators_style}
\footnotesize
\begin{tabular}{|c|c|}
\hline
\textbf{Accelerator Style}
& \textbf{Dataflow} \\
\hline
\multirow{3}{*}{FDA}
  & NVDLA \\
  \cline{2-2} 
  & Shi-diannao \\
  \cline{2-2}
  & Eyeriss \\
\hline
\multirow{3}{*}{Scaled-out Multi-FDA~\cite{baek2020multi}}
  & NVDLA + NVDLA  \\
    \cline{2-2} 
  & Shi-diannao + Shi-diannao \\
    \cline{2-2} 
  & Eyeriss + Eyeriss\\
\hline
RDA
& Flexible among three eval dataflows \\
\hline
\multirow{4}{*}{\shortstack[c]{HDA \\ (This work)}}
  & NVDLA + Shi-diannao (\textbf{Maelstrom}) \\
    \cline{2-2} 
  & Shi-diannao + Eyeriss \\
    \cline{2-2} 
  & Eyeriss + NVDLA \\
     \cline{2-2} 
  & NVDLA + Shi-diannao + Eyeriss \\
\hline
\end{tabular}
\end{table}

\noindent
\betterparagraph{Accelerator Styles}
For FDAs, we select NVDLA, Shi-diannao, and Eyeriss style accelerators.
For RDAs, we select MAERI~\cite{kwon2018maeri}.
We run MAESTRO to analyze latency and CAD tools with a 28nm library to analyze energy.
For HDAs, we select three two-way designs based on three dataflow styles selected for FDA and one three-way design combining all of those three dataflow styles.

We also model scale-out multi FDA (SM-FDA)~\cite{baek2020multi} that scales out an FDA architecture within an accelerator chip.
That is, sub-accelerators in an SM-FDA contain the same amount of hardware resources and run the same dataflow.
We apply \ourwork's scheduler to show the pure impact of homogeneous dataflow and evenly-partitioned hardware resources.

\begin{table}[]
\centering
\caption{Three accelerator classes for edge, mobile, and cloud use scenarios for evaluation.}
\label{tab:eval_accelerators_class}
\footnotesize
\begin{tabular}{|c|c|c|c|}
\hline
\textbf{Accelerator Class}
& \textbf{Num. of PEs} 
& \textbf{NoC BW} 
& \textbf{Glob. Memory} \\
\hline
Edge
& 1024
& 16 GB/s 
& 4 MiB \\
\hline
Mobile
& 4096 
& 64 GB/s 
& 8 MiB \\
\hline
Cloud
& 16384 
& 256 GB/s
& 16 MiB \\
\hline
\end{tabular}
\end{table}

\noindent
\betterparagraph{Accelerators Classes}
Based on previously proposed cloud and mobile accelerators ~\cite{jouppi2017datacenter,qualcomm_hexagon_680}, we select the amount of hardware resources for edge, mobile, and cloud use scenarios as described in~\autoref{tab:eval_accelerators_class}.

\betterparagraph{Schedulers}
We apply the scheduling algorithm we discussed in~\autoref{subsec:scheduling} in \ourwork.
We also implement a baseline greedy scheduler and compare our scheduler against it.

\subsection{Results}
\label{subsec:eval_results}


\insertWideFigure{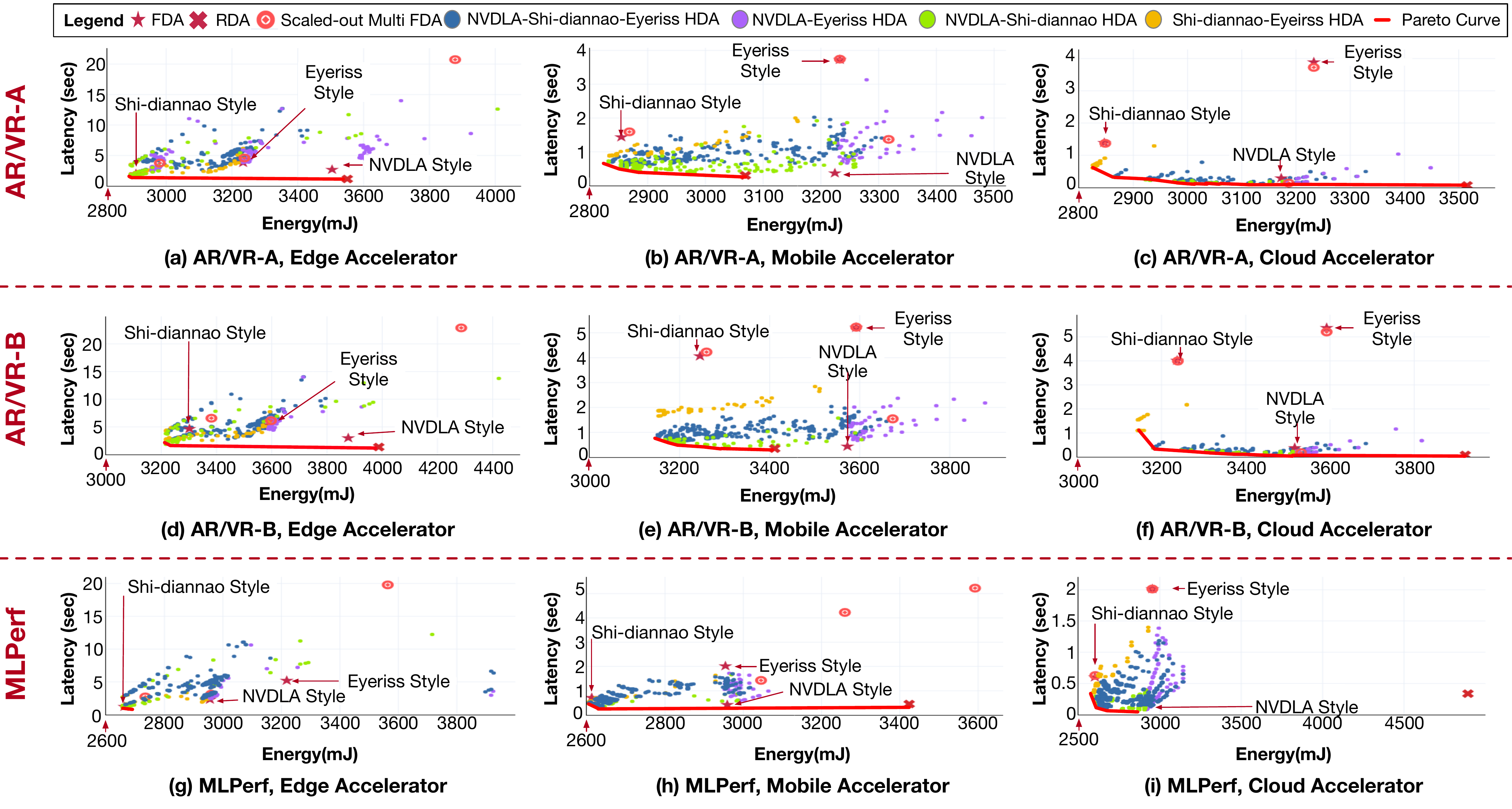}{Design space of two- and three-way HDAs.
Each row of the plots shows results for each workload listed in~\autoref{tab:eval_workloads} on three accelerator classes listed in~\autoref{tab:eval_accelerators_class}.
Each point in each plot represents a HW partitioning choice with an optimized schedule using \ourwork's scheduler. We label each FDA design point in each plot.}

We highlight some observations that provide useful insights from our evaluation.

\betterparagraph{Costs and Benefits of HDAs}
From~\autoref{fig:ThreewayDesignSpace}, we observe that well-optimized HDA and RDA design points are always on the Pareto curve over latency and energy, and FDA design points are not.
On average, compared to the best FDA design with the lowest EDP, the best heterogeneous design provided 73.6\% EDP improvements across all the case studies in~\autoref{fig:ThreewayDesignSpace}.

From the results in~\autoref{fig:ThreewayDesignSpace}, we identify that a two-way HDA architecture based on Shi-diannao and NVDLA dataflow styles provides the best latency and energy among four HDA designs we evaluate.
We name the HDA design with optimized hardware partitioning identified by \ourwork as \textbf{Maelstrom}.
We use Maelstrom as the reference HDA design for the rest of the evaluations.

\begin{table}[]
\centering
\caption{Maelstrom: Optimized HW Resource Partition found by \ourwork}
\label{tab:best_design_points}
\footnotesize
\begin{tabular}{|c|c|c|}
\hline
\multirow{2}{*}{\textbf{Scenario}}
& \textbf{BW Partitioning} 
& \textbf{PE Partitioning} 
\\ 
& \textbf{(NVDLA / Shi)}
& \textbf{(NVDLA / Shi)}
\\
\hline
AR/VR-A, Edge
& 4 / 12 
& 128 / 896 \\
\hline
AR/VR-A, Mobile
& 40 / 24
& 1792 / 2304 \\
\hline
AR/VR-A, Cloud
& 224 / 32
& 9728 / 6656 \\
\hline
AR/VR-B, Edge
&  4 / 12
&  128 / 896 \\
\hline
AR-VR-B, Mobile
& 48 / 16 
& 1536 / 2560 \\
\hline
AR/VR-B, Cloud
& 128 / 128 
& 12032 / 4352 \\
\hline
\rev{MLPerf}, Edge
&  \rev{4 / 12}
&  \rev{64 / 960} \\
\hline
\rev{MLPerf}, Mobile
&  \rev{32 / 32}
&  \rev{1280 / 2816} \\
\hline
\rev{MLPerf}, Cloud
& \rev{160 / 96}
& \rev{8192 / 8192} \\
\hline
\end{tabular}
\end{table}

Maelstrom demonstrates 65.30\% and 5.0\% lower latency and energy compared than the best FDA, 63.11\% and 4.1\% lower latency and energy than the SM-FDAs~\cite{baek2020multi}, and 20.7\% higher runtime but 22.0\% lower energy compared to a MAERI-based RDA~\cite{kwon2018maeri}.
Such benefits are based on the synergy of NVDLA and Shi-diannao dataflow styles.
When the number of channels is small, or the layer does not require accumulation across input channels (e.g., depth-wise CONV2D), NVDLA dataflow style significantly under-utilizes PEs.
Shi-diannao provides high efficiency on such layers because of its parallelization strategy across output activation rows and columns.
However, NVDLA provides higher efficiency for CONV2D layers with many channels and FC layers, which contains a large number of channels and performs accumulation across input channels.

\noindent
\betterparagraph{Optimal HW Resource Partitioning}
In~\autoref{tab:best_design_points}, we list the hardware resource partitioning results of Maelstrom design points with the best EDP for each workload and accelerator class.
We observe that the optimal hardware partitioning is not trivial (e.g., evenly partitioned), which necessitates a systematic approach like \ourwork.
This is because more number of active PEs requires more bandwidth, and the number of active PEs is a complex high-dimensional function of layer operation, layer size, number of PEs, mapping, and so on~\cite{kwon2019understanding}.

On average, across all the scenarios, we observe 111.12\% more PEs are assigned to NVDLA style, which implies more number of layers in the workloads prefer NVDLA style than Shi-diannao style.
However, on average, we observe 8.2\% more bandwidth is assigned to Shi-diannao style, which shows higher bandwidth requirements of Shi-diannao dataflow.

In particular, cloud accelerators have shown a stronger preference for NVDLA style dataflow, which resulted in 126.8\% and 59.3\% more bandwidth and PE assigned to NVDLA style sub-accelerator.
This is related to the degree of parallelism each dataflow can exploit from the layers in the workload.
NVDLA and Shi-diannao dataflows exploit channel and activation row/column parallelism, respectively.
The maximum channel parallelism in the workload is 16.8M (FC layer 2, Focal Length DepthNet~\cite{he2018learning}), but the maximum activation parallelism in the workload is 334.1K (CONV layer 1, UNet~\cite{UNet}), which led to a stronger preference to NVDLA dataflow-style dataflow that exploits channel-parallelism.
The maximum parallelization degree implies that we can design more powerful and efficient Maelstrom up to 16.8M PEs for three evaluated workloads until other conditions (e.g., memory, chip area, power, etc.) allow.

Although the workload is overall friendlier to NVDLA, the best designs of NVDLA-Shi-diannao HDAs, Maelstrom, balances between NVDLA and Shi-diannao styles, as shown in~\autoref{tab:best_design_points}.
Such optimization results imply that the workload is highly heterogeneous, and Maelstrom successfully exploited its dataflow heterogeneity for the heterogeneous workload.

\noindent
\betterparagraph{Impact of Workloads}
Each row in~\autoref{fig:ThreewayDesignSpace} shows the design space of three evaluation workloads, where we can observe the design space of HDAs depends on the workloads.
In particular, we observe that workload with more heterogeneity and layers like AR/VR-B workload is more friendly to HDAs, providing 86.8\% latency and 6.61\% energy improvements over bestFDAs for each case study in~\autoref{fig:ThreewayDesignSpace}, compared to 63.26\% latency and 4.05\% energy improvements for AR/VR-A and \rev{48.1\%} latency and \rev{4.4\%} energy improvements for MLPerf.

\insertFigure{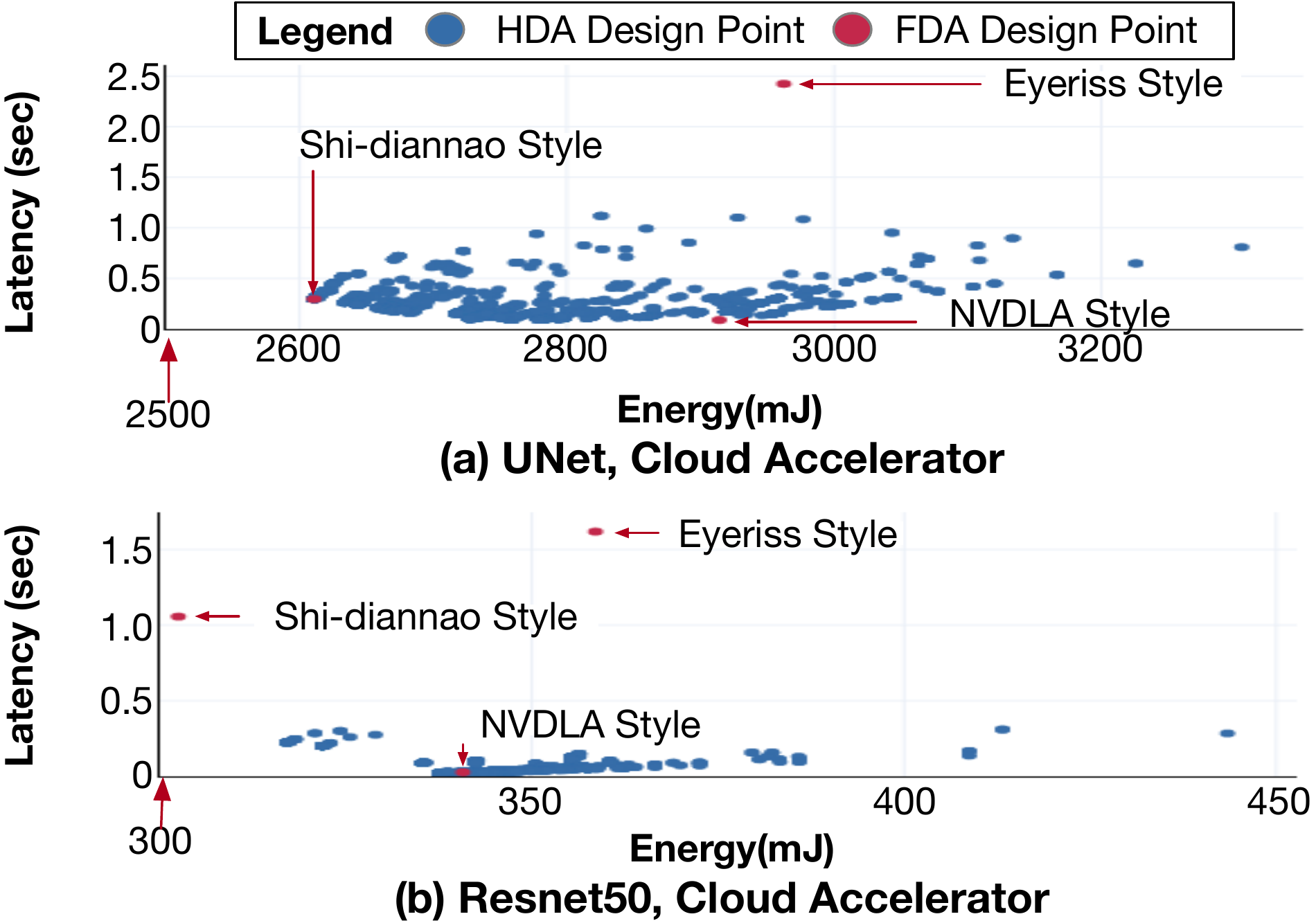}{Design space of single DNN use cases on (a) UNet and (b) Resnet50 based on cloud accelerator settings in~\autoref{tab:eval_accelerators_class}.}

\betterparagraph{Single-DNN Case}
Even for a single DNN, HDAs can still exploit layer parallelism and heterogeneity within a model by batch-processing the workload.
We run UNet and Resnet50 using the batch size of four on the cloud scenario and \rev{present the results in~\autoref{fig:SingleDNN}}.
We observe that the best FDA design is on the Pareto curve, unlike heterogeneous multi-DNN workloads we target.
However, optimized Maelstrom designs still provide latency and energy benefits over monolithic designs.
For UNet and Resnet50 workload, Maelstrom provided 26.4\% and 48.1\% EDP improvements over the best monolithic design.
RDAs provided 22.5\% and 29.0\% lower latency compared to Maelstrom for UNet and Resnet50, respectively.
However, RDAs required 11.7\% and 15.8\% more energy than Maelstrom for UNet and Resnet50, respectively.

\rev{
\betterparagraph{Efficacy of Scheduling Algorithm}
}
We compare the EDP of schedules from \ourwork's scheduler and \rev{a greedy scheduler, that assigns a sub-accelerator with the least EDP for each layer.}.
%
%
\rev{
Compared to the greedy scheduler, \ourwork's scheduler considers global load balancing, exploits the layer dependence chain, and performs the post-processing discussed in~\autoref{fig:Scheduling_Overview}.  
On average, \ourwork's scheduler identified schedules on Maelstrom with 24.1\% less EDP, compared to the greedy scheduler}
%


\begin{table}[]
\centering
\caption{\rev{Latency and energy gain against the FDA and RDA with the best EDP on various batch sizes on MLPerf workload.}}
\label{tab:eval_batchsize}
\footnotesize
\begin{tabular}{|c|c|c|c|}
\hline
\multirow{2}{*}{\textbf{Acc. Class}}
& \multirow{2}{*}{\textbf{Batch Size}}
& \textbf{Latency Gain}
& \textbf{Energy Gain} \\ 
&
& \textbf{(vs FDA / vs RDA)}
& \textbf{(vs FDA / vs RDA)} \\ 
\hline
\multirow{2}{*}{Edge}
& 1 & \rev{12.4\% / -8.2\%} & \rev{0.2\% / 20.4\%} \\
\cline{2-4}
& 8 & \rev{21.28\% / 26.7\%} & \rev{10.8\% / 22.9\%} \\
\hline
\multirow{2}{*}{Mobile}
& 1 & \rev{12.4\% / -8.2\%} & \rev{0.2\% / 17.1\%}\\
\cline{2-4}
& 8 & \rev{56.0\% / 76.1\%} & \rev{1.3\% / 43.5\%} \\
\hline
\multirow{2}{*}{Cloud}
& 1 & \rev{20.2\% / 25.7\%} & \rev{10.8\% / 26.8\%} \\
\cline{2-4}
& 8 & \rev{63.9\% / 80.4\%} & \rev{1.34\% / 41.3\%} \\
\hline
\end{tabular}
\end{table}

\rev{\betterparagraph{Impact of Batch Size}}
\rev{We vary the batch size of MLPerf workload from one to eight and quantify the latency and energy benefits of HDAs.
We summarize the latency and energy gain of HDAs in~\autoref{tab:eval_batchsize}.
We observe that HDA prefers large batch sizes, and HDA can outperform RDAs in both latency and energy when the batch size is large.
On average, compared to RDAs, HDAs provided 3.1\% latency and 21.4\% energy savings on MLperf workload with batch size 1.
When the batch size is increased to 8, HDAs provided 61.1\% less latency and 35.9\% less energy compared to RDAs, which shows HDA's preference for large batch sizes.
}

\betterparagraph{Comparison against RDAs}
We evaluate a MAERI~\cite{kwon2018maeri} style RDA and present the design points in~\autoref{fig:ThreewayDesignSpace}.
Compared to Maelstrom in each scenario, RDA designs provided 22.9\%, 21.5\%, and \rev{24.3\%} less latency for AR/VR-A, AR/VR-B, and \rev{MLPerf} workloads, respectively.
However, RDA designs required 18.7\%, 15.5\%, and \rev{18.9\%} more energy for each workload, respectively.
The extra energy cost of RDA is based on hardware components for reconfigurability.
In contrast, an HDA can keep sub-accelerators with relatively simple architecture compared to flexible accelerators, which leads to better energy efficiency we present.

Results in~\autoref{fig:ThreewayDesignSpace} show that both HDA and RDA architectures are Pareto-optimal.
HDAs and RDAs have strength in energy and latency, respectively.
The amount of benefits for latency and energy depends on the workload.
Therefore, the choice of RDA or HDA depends on the performance goal, energy constraints, and the target workload.

\insertFigure{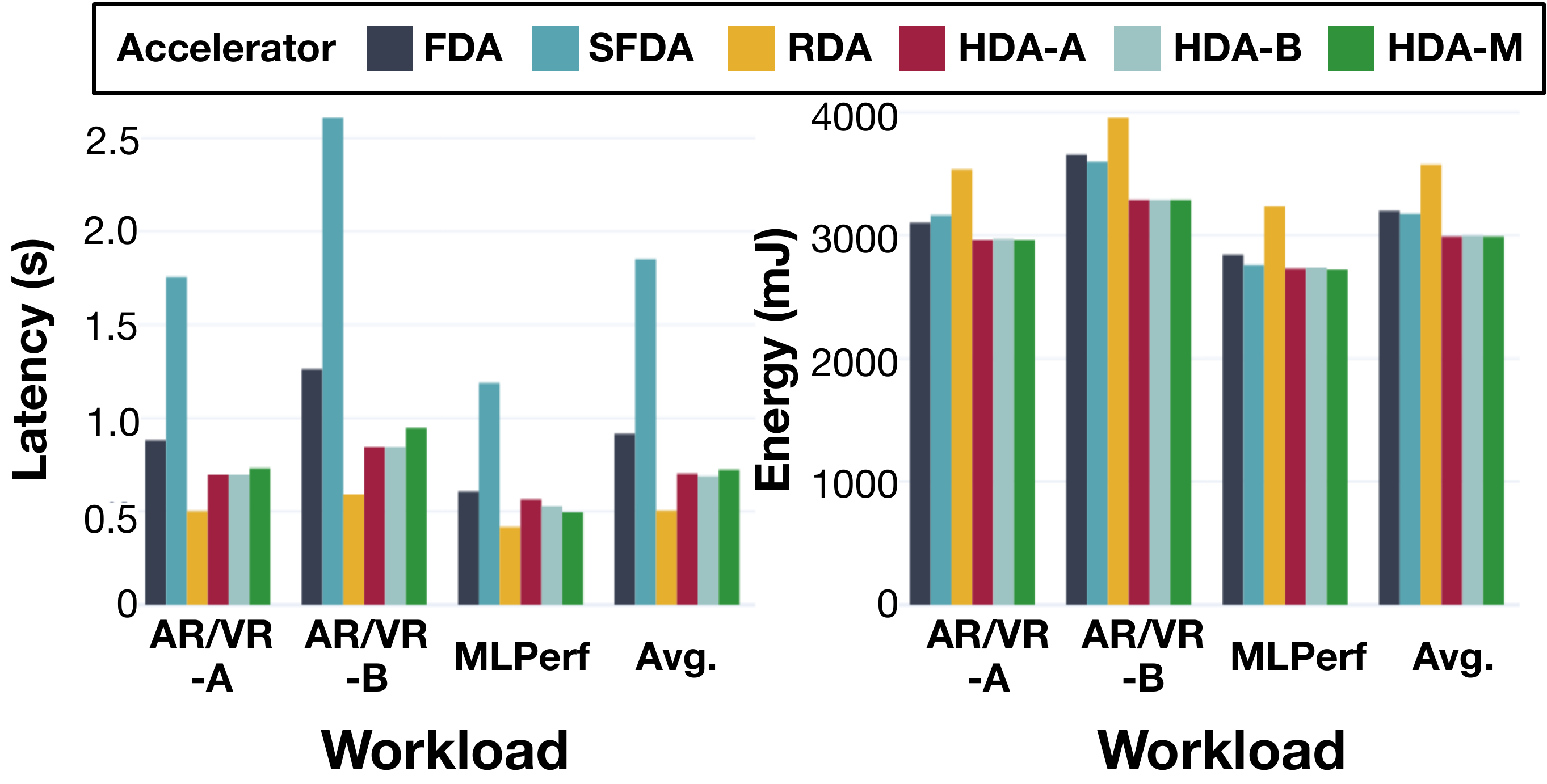}{\rev{ Average latency and energy across edge, mobile, and cloud accelerator classes for each workload. HDA-A, HDA-B, and HDA-M refer to Maelstrom designs optimized for AR/VR-A, AR/VR-B, and MLPerf workload, respectively. SFDA refers to scaled-out FDA.}}

\noindent
\rev{\betterparagraph{Impact of Workload Change}
Since DNN models evolve and applications change their inner implementation accordingly, workload change can occur after the deployment of an HDA.
To deep-dive into the impact of such workload change after deployment of HDAs, we perform a case study of workload change in~\autoref{fig:WorkloadChange_Study}.
In the case study, we fix each HDA design and only perform layer execution scheduling for running different workloads than a workload each HDA design is optimized for (e.g., running AR/VR-B workload on an HDA originally optimized for AR/VR-A workload).
}

\rev{
From the case study, we observe that running different workloads than the workload each HDA is optimized for results in minor latency and energy increase, 4.0\% and 0.1\%, on average. 
On average, across all the workload change scenarios, HDAs provided 19.44\% energy savings at the cost of 28.6\% latency against RDAs.
Against FDAs, HDAs provided 29.99\% latency and 6.45\% energy savings, on average.
}

%
%
%

\begin{table}[]
\centering
\caption{Average time required for scheduling each workload on HDAs.}
\label{tab:eval_exec_time}
\footnotesize
\begin{tabular}{|c|c|c|c|}
\hline
\textbf{Workload}
& \textbf{\# Layers}
& \textbf{\# sub-accelerators}
& \textbf{Scheduling Time (s)} \\ 
\hline
\multirow{2}{*}{AR/VR-A} &
\multirow{2}{*}{448} &
2 & 2.89  \\
\cline{3-4}
& & 3 & 4.32 \\
\hline
\multirow{2}{*}{AR/VR-B} &
\multirow{2}{*}{618} &
2 & 3.98  \\
\cline{3-4}
& & 3 & 10.74 \\
\hline
\multirow{2}{*}{\rev{MLPerf}} &
\multirow{2}{*}{\rev{181}} &
2 & \rev{1.61}  \\
\cline{3-4}
& & 3 & \rev{3.22} \\
\hline
\end{tabular}
\end{table}

\betterparagraph{Scheduling Time}
Although \ourwork's scheduler is designed to be offline, the scheduler is light-weighted.
We run \ourwork on a laptop with i9-9880H processor and 16GB memory and present the time required for scheduling on each HDA design point in~\autoref{tab:eval_exec_time}, since overall execution time heavily depends on user parameters (e.g., search granularity, number of sub-accelerators, etc.).
On average, the scheduling requires \rev{11.09 ms} per layer and per HDA design point.

%

\noindent
\betterparagraph{Summary}
We summarize our main observations below:

\squishlist
    {\item The design space of HDA is not trivial, which requires a systematic co-optimization of hardware resource partitioning and layer execution schedule.}
    {\item We identify a promising HDA architecture Maelstrom based on NVDLA and Shi-diannao dataflow styles. Maelstrom designs outperform FDA and SM-FDA designs in overall latency and energy.}
    {\item Both Maelstrom and RDA architectures are Pareto-optimal design points with different strengths in energy efficiency and latency, respectively.
    }
    {\item Simple combination of sub-accelerators running the same dataflow does not lead to Pareto-optimal design points, which shows the efficacy of HDAs.}
\squishend

\section{Related Works}
\label{sec:related_works}

%

\betterparagraph{DNN Dataflows and Accelerators} 
Shi-diannao~\cite{du2015shidiannao} is an FDA designed to be embedded near sensors, which exploits convolutional reuse via an output-stationary style dataflow.
Eyeriss~\cite{chen2016eyeriss_jssc} is one of the state-of-the-art low-energy DNN accelerators that introduced dataflow taxonomy and a new dataflow style, row-stationary.
Fused-layer CNN accelerator~\cite{alwani2016fused} exploited fine-grained pipelined layer parallelism that minimizes activation data movement.
Flexflow~\cite{lu2017flexflow} is an RDA that supports three distinct dataflows.
Tensor Processing Unit(TPU)~\cite{jouppi2017datacenter} is a systolic array-based DNN accelerator designed for cloud workload in data centers. 
MAERI~\cite{kwon2018maeri} is an RDA that efficiently supports irregular mappings resulting from sparsity, cross-layer mapping~\cite{alwani2016fused}, and so on. 
Tangram~\cite{gao2019tangram} is a DNN accelerator that explored pipelined layer parallelism within a model with optimized dataflow for such back-to-back layer execution.
Interstellar~\cite{yang2020interstellar} presented the importance of loop blocking (tile sizing) in DNN accelerators utilizing Halide~\cite{ragan2013halide}.
Prema~\cite{choi2020prema} explored the use of preemption scheduler implemented in hardware for QoS of multiple-DNNs, which targets a systolic array-based FDA.

\betterparagraph{Multi-DNN Accelerators}
Shen et al. explored the use of multiple FDA sub-accelerators running the same dataflow termed as convolutional layer processors in FPGAs~\cite{shen2017maximizing}.
AI-multitasking architecture~\cite{baek2020multi} employed multiple systolic arrays within an accelerator chip and parallelized computation tiles of each layer.
Although the idea of employing sub-accelerators is proposed in those works, \textit{\ourwork first explored the dataflow heterogeneity using sub-accelerators, co-optimizing hardware resource partitioning optimization and layer execution schedule.}

\betterparagraph{Heterogeneous Accelerators}
Chandramoorthy et al.~\cite{chandramoorthy2015exploring} explored accelerator-rich chip-multiprocessor that include various accelerators for different tasks in image processing.
Although the work included a convolution module among sub-accelerators, the convolution module provides only one dataflow style, focusing on general image kernels, not DNNs. 
Master of None Acceleration~\cite{lottarini2019master} explored a heterogeneous accelerator for analytical query and presented that the design space of heterogeneous accelerators for the target domain has both beneficial and disadvantageous design points.

\section{Conclusion}
\label{sec:conclusion}

In this paper, we explored the latency and energy optimization opportunities of heterogeneous dataflow accelerators (HDAs) on recent heterogeneous multi-DNN workloads such as AR/VR.
Because the efficiency of a DNN accelerator depends on mapping, workload, and hardware design parameters at the same time, identifying the best HDA design point with an optimized schedule is challenging.
Therefore, we developed \ourwork, an automated co-design space exploration framework for hardware resource partitioning and layer scheduling for heterogeneous DNN accelerators.
In our evaluation, \ourwork identified a promising HDA architecture, Maelstrom, which deploys NVDLA and Shi-diannao dataflow styles over two sub-accelerators.
Maelstrom provided, on average, 73.6\% EDP benefits compared to the best fixed dataflow accelerator designs we compare across three workloads we evaluate.
We observe that the most efficient Maelstrom design points have non-trivial hardware resource partitioning, and a naive scheduler can result in EDP degradation, motivating a systematic approach like \ourwork.

In summary, HDA is a new promising class of flexible dataflow accelerators, and \ourwork facilitates the design of HDAs via co-optimization of hardware resource partitioning across sub-accelerators and layer execution schedule.

\section*{Acknowledgements}
We thank Simon Hollis, Meng Li, Pierce Chuang, Ganesh Venkatesh, and Yilei Li for insightful comments and discussions.
This work was supported in-part by NSF Award OAC-1909900.


\bibliographystyle{IEEEtran}
\bibliography{ref}

\end{document}